\documentclass[11pt,showpacs,preprintnumbers,amsmath,amssymb,prd,nofootinbib,superscriptaddress]{revtex4-2}

\usepackage{dcolumn}
\usepackage{bm}
\usepackage{ifpdf}
\usepackage{hyperref}
\usepackage{bm}
\usepackage{xcolor,color,graphicx,graphics}
\usepackage[spanish,english]{babel}
\usepackage[latin1]{inputenc}
\usepackage[OT1]{fontenc}
\usepackage{latexsym,amssymb,amsmath,amsfonts}
\usepackage{makeidx}
\usepackage{epsfig,subfigure}
\usepackage{natbib}
\usepackage{epstopdf}
\usepackage{mathrsfs}
\usepackage{hyperref}
\hypersetup{colorlinks=true, linkcolor=blue, citecolor=green}
\usepackage{enumerate}

\everymath{\displaystyle}
\usepackage{graphicx}

\usepackage[T1]{fontenc}
\usepackage{amsmath}
\usepackage{amssymb}
\usepackage{graphicx}
\usepackage{xcolor}

\newcommand{\bea}{\begin{eqnarray}}
\newcommand{\eea}{\end{eqnarray}}

\newcommand{\orcid}[1]{\href{https://orcid.org/#1}{\includegraphics[width=10pt]{orcid}}}

\usepackage{fixmath}

\newcommand{\sumspin}{\sum_{\mathtt{\,\,\,Spin}}\!}

\begin{document}

\title{Gravitational electron-positron scattering}

\author{W. D. R. Jesus}
\email{willian.xb@gmail.com}
\affiliation{Instituto Federal de Educação, Ci\^encia e Tecnologia de Mato Grosso,\\
78043-409, Cuiab\'{a}, Mato Grosso, Brazil}

\author{P. R. A. Souza}
\email{pablo@fisica.ufmt.br}
\affiliation{Instituto de F\'{\i}sica, Universidade Federal de Mato Grosso,\\
78060-900, Cuiab\'{a}, Mato Grosso, Brazil}

\author{A. F. Santos} 
\email{alesandroferreira@fisica.ufmt.br}
\affiliation{Instituto de F\'{\i}sica, Universidade Federal de Mato Grosso,\\
78060-900, Cuiab\'{a}, Mato Grosso, Brazil}

\author{Faqir C. Khanna \footnote{Professor Emeritus - Physics Department, Theoretical Physics Institute, University of Alberta\\
Edmonton, Alberta, Canada}}
\email{fkhanna@ualberta.ca; khannaf@uvic.ca}
\affiliation{Department of Physics and Astronomy, University of Victoria,\\
3800 Finnerty Road, Victoria BC V8P 5C2, Canada}

\begin{abstract}

A scattering process with gravitons as an intermediate state is investigated. To study such a scattering, the Gravitoelectromagnetism theory is considered. It is a gravitational theory built on the analogy between gravity and electromagnetism. The complete Lagrangian formulation of the gravitoelectromagnetic theory includes interactions of gravitons with fermions and photons that leads us to calculate their scattering amplitudes and cross-sections. In this context, the gravitational cross-section of the $e^{-}+e^{+}\longrightarrow\mu^{-}+\mu^{+}$ scattering process is obtained. A comparison between the electromagnetic and gravitational cross-sections is made. 


\end{abstract}

\maketitle

\section{Introduction}

The interactions between elementary particles are described by the standard model which is a gauge theory, based on the group $\mathrm{U(1)\times SU(2)\times SU(3)}$. The standard model describes strong, weak, and electromagnetic interactions \cite{Book1, Book2} and since its formulation, it has been extensively tested with great success. However, gravitational interaction is not included in the standard model. The theory of gravity is classically described by the theory of general relativity. Then a natural question that arises is: how are interactions between gravitons and other elementary particles investigated? In order to answer this question, the theory of  Gravitoelectromagnetism (GEM) is considered.

The GEM theory is an old idea that emerges from attempts to unify the theories of gravity and electromagnetism \cite{Faraday, Maxwell, Heaviside1, Heaviside2, Weyl, KK}. The foundation of this theory lies in the similarities between gravitational and electromagnetic forces. There are three different approaches to investigating GEM theory: (i) considering the Einstein and Maxwell equations in the weak field approach \cite{Mashhon}; (ii) using tidal tensors \cite{Filipe} and (iii) using the Weyl tensor and its decomposition into components \cite{Maartens}. Weyl components are designated as gravitoelectric and gravitomagnetic fields. These names are associated with the fact that in weak field approximation the Einstein equations take a form similar to Maxwell equations of electromagnetism. Thus, the interaction between moving masses can be described in terms of forces due to two fields: the gravitoelectric which is analogous to the electric field, and the gravitomagnetic field which is analogous to the magnetic field of classical electromagnetism. Some effects of the GEM theory have been experimentally confirmed \cite{Large, Probe}.

In this paper, the Weyl tensor approach is considered. In this context, there is a certain correspondence between the Weyl tensor $C_{\alpha\sigma\mu\nu}$ which represents the free gravitational field and the electromagnetic tensor $F_{\alpha\sigma}$ which is associated with the electromagnetic phenomena. However, this analogy is not general, there are some limits. An important example is the fact that the gravitational field itself generates the space-time while the electromagnetic field propagates in a given space-time, among other limitations.

A Lagrangian formulation for the GEM theory has been constructed \cite{Khanna}. This gravitational Lagrangian allow us to consider the gravitons and their interactions with other elementary particles such as fermions and photons. From this formalism, some studies have been developed. For example, the gravitational Bhabha scattering has been investigated \cite{BhabhaGEM, BhabhaGEMTem, BhabhaGEMTemLV}, gravitational M\"{o}ller scattering has been analyzed \cite{Moller, Moller1}, the gavitational Casimir effect at finite temperature has been calculated \cite{Casimir}, among others.
Here the gravitational electron-positron scattering is investigated.

The $e^{-}+e^{+}\longrightarrow\mu^{-}+\mu^{+}$ scattering process is a well-known quantum electrodynamics process. It occurs via the annihilation of electron-positron into virtual photon, and then the creation of a muon pair. In this paper, a gravitational version of this process is considered. The GEM Lagrangian formalism allows us to investigate gravitational contribution to the usual electromagnetic scattering processes. In this case,  there is a graviton as an intermediate state, in addition to a photon as an intermediate state.

This paper is organized as follows. In section II, an introduction to the GEM theory is developed. In the weak-field approximation, the equivalence between GEM theory and general relativity is shown. In the section III, the scattering process and the centre of mass reference frame are presented. The interaction part of the Lagrangian and the vertex are introduced. In section IV, the transition amplitude and the gravitational cross section are calculated. The electromagnetic and gravitational cross sections of electron-positron scattering are compared. In section V, some concluding remarks are presented.

\section{GEM theory}

The main objective of this section is to obtain a Lagrangian formulation for this theory of gravity. This construction is developed considering the gravitational (GEM) or Maxwell-like equations that are given as
\bea
&&\partial^i{\cal E}^{ij}=-4\pi G\rho^j,\label{01}\\
&&\partial^i{\cal B}^{ij}=0,\label{02}\\
&&\epsilon^{( i|kl}\partial^k{\cal B}^{l|j)}+\frac{\partial{\cal E}^{ij}}{\partial t}=-4\pi G J^{ij},\label{03}\\
&&\epsilon^{( i|kl}\partial^k{\cal E}^{l|j)}+\frac{\partial{\cal B}^{ij}}{\partial t}=0,\label{04}
\eea
with ${\cal E}^{ij}$ and ${\cal B}^{ij}$ being the gravitoelectric and gravitomagnetic fields, respectively, $G$ being the gravitational constant,  $\rho^j$ the vector mass density and $J^{ij}$ the mass current density. The GEM fields constructed from the Weyl tensor and its dual are written as
\bea
{\cal E}_{ab}=C_{abcd}u^cu^d, \quad\quad\quad {\cal B}_{ab}=\frac{1}{2}\epsilon_{acd}C^{cd}\,_{be}u^e,
\eea
where $u^a$ is a 4-velocity vector field and 
\bea
C_{\alpha\sigma\mu\nu}&=&R_{\alpha\sigma\mu\nu}-\frac{1}{2}\left(R_{\nu\alpha}g_{\mu\sigma}+R_{\mu\sigma}g_{\nu\alpha}-R_{\nu\sigma}g_{\mu\alpha}-R_{\mu\alpha}g_{\nu\sigma}\right)+\frac{1}{6}R\left(g_{\nu\alpha}g_{\mu\sigma}-g_{\nu\sigma}g_{\mu\alpha}\right)
\eea
is the Weyl tensor. Here $R_{\alpha\sigma\mu\nu}$, $R_{\mu\nu}$ and $R$  are the Riemann tensor, the Ricci tensor and the Ricci scalar, respectively.

In order to obtain the GEM Lagrangian \cite{Khanna}, a gravitoelectromagnetic tensor potential ${\cal A}^{\mu\nu}$ is considered. Then the GEM fields are written as
\bea
{\cal E}&=&-\mathrm{grad}\,\varphi-\frac{\partial \tilde{\cal A}}{\partial t},\\
{\cal B}&=&\mathrm{curl}\,\tilde{\cal A},
\eea
where $\varphi$ is the GEM counterpart of the electromagnetic scalar potential $\phi$. Thus the GEM equations (\ref{01})-(\ref{04}) become
\bea
\partial_\mu{ F}^{\mu\nu\alpha}&=&4\pi G{\cal J}^{\nu\alpha},\label{FE}\\
\partial_\mu{\cal G}^{\mu\langle\nu\alpha\rangle}&=&0,
\eea
with ${\cal J}^{\nu\alpha}$ being a tensor related to $\rho^i$ and $J^{ij}$. In addition, the GEM tensor ${ F}^{\mu\nu\alpha}$ and the GEM dual tensor ${\cal G}^{\mu\nu\alpha}$ are defined, respectively, as 
\bea
{F}^{\mu\nu\alpha}&=&\partial^\mu{\cal A}^{\nu\alpha}-\partial^\nu{\cal A}^{\mu\alpha},\\\label{Tensor}
{\cal G}^{\mu\nu\alpha}&=&\frac{1}{2}\epsilon^{\mu\nu\gamma\sigma}\eta^{\alpha\beta}{F}_{\gamma\sigma\beta}.
\eea
It should be noted that ${F}^{0ij}={\cal E}^{ij}$ and ${F}^{ijk}=\epsilon^{ijl}{\cal B}^{lk}$, where $i, j=1, 2, 3$.

Therefore, using these ingredients, the GEM Lagrangian density \cite{Khanna} is given as
\bea
{\cal L}_{GEM}=-\frac{1}{16\pi}{F}_{\mu\nu\alpha}{F}^{\mu\nu\alpha}-G\,{\cal J}^{\nu\alpha}{\cal A}_{\nu\alpha}.\label{L_G}
\eea

It is interesting to note that in the weak field approximation the GEM and general relativity are equivalent. To verify such equivalence, eq. (\ref{FE}) is written as
\bea
\Box A^{\nu\alpha}-\partial^\nu(\partial_\mu A^{\mu\alpha})=4\pi G{\cal J}^{\nu\alpha}.\label{12}
\eea
This equation in the Lorenz-like gauge, i.e. $\partial_\mu A^{\mu\alpha}=0$, reads
\bea
\Box A^{\nu\alpha}=4\pi G{\cal J}^{\nu\alpha}.\label{GEM}
\eea
Now let us consider the general relativity in the weak-field approximation. This implies that
\bea
g_{\mu\nu}=\eta_{\mu\nu}+\kappa h_{\mu\nu},
\eea
where $h_{\mu\nu}$ is a small perturbation and $\kappa^2=32\pi G$. Then the Einstein equations become
\bea
\Box\bar{h}_{\mu\nu}=16\pi G\,T_{\mu\nu}\label{GR}
\eea
with $T_{\mu\nu}$ being the energy-momentum tensor and $\bar{h}^{\mu\nu},_\nu=0$ represents the condition of the harmonic gauge. Thus, eqs. (\ref{GEM}) and (\ref{GR}) imply that GEM theory and  Einstein linearized equations are equivalent. However, there is an essential difference, the GEM potential,  $A_{\mu\nu}$, is a fundamental gravitational field and $h_{\mu\nu}$ is a small perturbation. In addition, there is no relationship between these two quantities.

In order to calculate the cross-section for electron-positron scattering in the GEM field, the complete Lagrangian that describes the interaction between gravitons and fermions is provided in the next section.

\section{The Electron-positron scattering}

In this section, the $e^{-}+e^{+}\longrightarrow\mu^{-}+\mu^{+}$ scattering process with gravitons in the intermediate state is investigated. The Feynman diagram that describes this process is shown in Figure 1.
\begin{figure}[h]
\centering{}
\includegraphics[clip,scale=0.47]{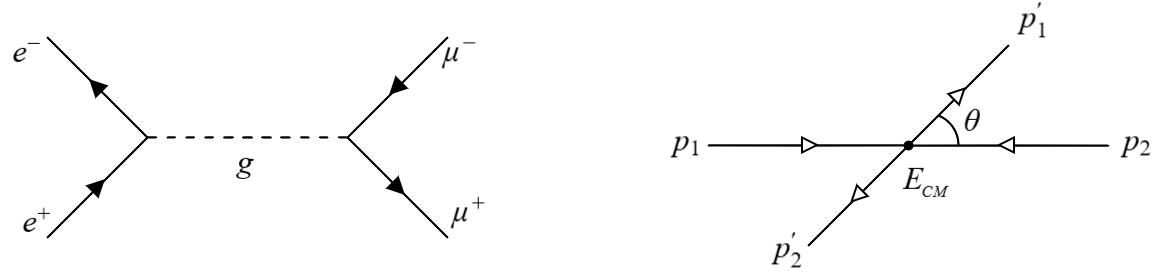}
\caption{\footnotesize{On the left, the Feynman diagram that describes the scattering process. On the right, the process is shown in the center of mass (CM) frame.}}
\end{figure}

Here the scattering process is considered in the centre of mass (CM) reference frame, where
\begin{eqnarray}
&& p_1=E(1,0,0,1);\quad\quad\quad\quad\quad\quad p_2=E(1,0,0,-1);\nonumber\\
&& p'_1=E(1,\sin(\theta),0,\cos(\theta));\quad\quad p'_2=E(1,-\sin(\theta),0,-\cos(\theta)),\
\end{eqnarray}
with $p_i$ and $p'_i$ ($i,j=1,2$) being the momentum associated to the electron (positron) and muon (anti-muon), respectively. In addition, $p_1\equiv \mathrm{p}_{1\mu}=(p_0, \textbf{p})$, $p'_1\equiv \mathrm{p}'_{1\mu}=(p_0, -\textbf{p})$ and $|\kappa_0|=|p_2-p_1|=|p'_2-p'_1|=\mathrm{E_{_{CM}}=s=4E^2}$.

The GEM Lagrangian that describes this scattering process is given by
\begin{equation}
\mathcal{L}_{_\mathrm{GEM}}= -\frac{1}{16\pi}F_{\mu\nu\alpha}F^{\mu\nu\alpha}-\frac{1}{2}\Big(\overline{\Psi}\gamma^\mu\partial_\mu\Psi-(\partial^\mu\overline{\Psi})\gamma_\mu\mu\Psi\Big)+ m^2\Psi\overline{\Psi}
+\mathcal{L}_{_{(I)\mathrm{GEM}}},
\end{equation}
where the interaction part is
\begin{equation}
\mathcal{L}_{_{(I)\mathrm{GEM}}}=- \imath\frac{g}{4}A_{\mu\nu}\Big(\overline{\Psi}\gamma^\mu\partial^\nu\Psi-(\partial^\mu\overline{\Psi})\gamma^\nu\Psi\Big)\label{interaction}
\end{equation}
with $\Psi$ being the fermion field, $m$ the fermion mass, $\gamma^\mu$ the Dirac matrices and $g=\frac{\sqrt{8\pi G}}{c^2}$ the coupling constant. 

The graviton propagator $D_{\mu\nu\alpha\beta}(\kappa)$ and the graviton-fermions vertex $\Gamma^{\mu\nu}$ are defined, respectively, as
\bea
D_{\mu\nu\alpha\beta}(\kappa)&=&\frac{i}{2\kappa^2}\left(\eta_{\mu\alpha}\eta_{\nu\beta}+ \eta_{\mu\beta}\eta_{\nu\alpha} - \eta_{\mu\nu}\eta_{\alpha\beta}\right),\label{prop}\\
\Gamma^{\mu\nu}&=&-\frac{ig}{4}\left(\gamma^\mu p_1^\nu+p_2^\mu\gamma^\nu\right),
\eea
with $\kappa$ being the momentum transferred. Feynman diagrams representing these quantities are given in Figure 2.
\begin{figure}[h]
\centering{}
\includegraphics[clip,scale=0.6]{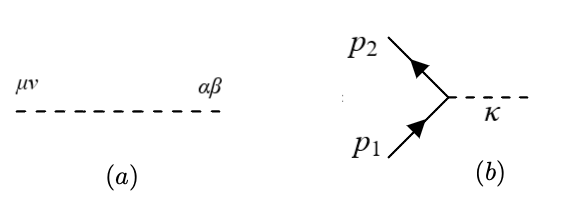}
\caption{The graviton propagator and vertex are shown in (a) and (b), respectively.}
\end{figure} 

In the next section, the transition amplitude is obtained. Then the gravitational cross-section of the electron-positron scattering is calculated.

\section{Gravitational Cross-section}

To obtain the cross-section, the main quantity that must be calculated is the transition amplitude $\mathcal{M}$ which is defined as 
\begin{equation}\label{M0}
 \mathcal{M}=\Big\langle f\Big|{S}^{(2)}\Big|\imath\Big\rangle=
\mathrm{\frac{(-\imath)^2}{2}\int\!dx\,dy\Big\langle f\,\Big|:\mathcal{L}_{_{(I)\mathrm{GEM}}}(x)\mathcal{L}_{_{(I)\mathrm{GEM}}}(y):\Big|\,\imath\Big\rangle}
\end{equation}
where the $S$-matrix second-order term is given as 
\begin{equation}
{S}^{(2)}
=\mathrm{\frac{(-\imath)^2}{2}\!\int\!\!dx\,dy\big[:\mathcal{L}_{_{(I)\mathrm{GEM}}}(x)\mathcal{L}_{_{(I)\mathrm{GEM}}}(y):\big]}.
\end{equation}
Here the notation $``:\,\,\,:"$ represents the time ordering operation. The initial and final asymptotic states are
\begin{equation}\label{estados0}
      \Big|\,\imath\Big\rangle = c^\dagger_{p_i}(\mathtt{s}_i)d^\dagger_{p_j}(\mathtt{s}_j)\Big|\, 0\Big\rangle\quad \mbox{and}\quad \Big|\,f\Big\rangle = c^\dagger_{p_k}(\mathtt{s}_k)d^\dagger_{p_l}(\mathtt{s}_l)\Big|\, 0\Big\rangle,
\end{equation}
where $\mathtt{s}_i$ are the spin variables, $c^\dagger_q$ $(d^\dagger_q)$ and $c_q$ $(d_q)$ are the creation and annihilation operators, respectively. These operators satisfy the anti-commutation relations 
\begin{equation}\label{comutacao0}
\{c_{p_i}(\mathtt{s}_i);\, c^\dagger_{q_j}(\sigma_j) \} =\delta_{\mathtt{s}_i\sigma_j}\delta^{(3)}(p_i-q_j)\quad \mbox{and}\quad \{d_{p_i}(\mathtt{s}_i);\, d^\dagger_{q_j}(\sigma_j) \} =\delta_{\mathtt{s}_i\sigma_j}\delta^{(3)}(p_i-q_j),
\end{equation}
and other anti-commutation relations are null.

Using these quantities, the differential cross-section is calculated. It is defined as
\begin{equation}\label{sigma0}
   \frac{d\sigma}{d\Omega}
   =\frac{|\textbf{p}'|}{64\pi^2\,s|\textbf{p}|}\Big\langle\,\,\Big|\mathcal{M}\Big|^2\,\Big\rangle\,,
\end{equation}
where $|\textbf{p}'|$ and $|\textbf{p}|$ are the final and initial moments, respectively, and
\begin{equation}\label{Mmedio0}
\Big\langle\,\,\Big|\mathcal{M}\,\Big|^2\,\Big\rangle=\frac{1}{4}\sum_{\mathrm{Spins}}\Big|\mathcal{M}\Big|^2\,.
\end{equation}
Here, an average over the spin of the incoming particles and summing over the spin of the outgoing particles is assumed.
 
In order to calculate the transition amplitude, let us consider the fermion field given as
\begin{equation}\label{psi}
\Psi(x)=\int\!\! dp_i\Big[c_{p_i}u(i)e^{ix\cdot p_i}+d^{\dagger}_{p_i}\overline{u}(i)e^{-ix\cdot p_i}\Big],
\end{equation}
where $u(i)\equiv u(p_i, s_i)$ has been used with $i=1,2,3,4$. Using eq. (\ref{interaction}), the transition amplitude eq. (\ref{M0}) becomes
\begin{equation}\label{amplitudePAA}
\mathcal{M}=\frac{(-\imath)^2}{2}\!\int\!\!\int\!\mathrm{dxdy}\Big\langle f\Big|:\mathfrak{P}^{\mu\nu\alpha\beta}A_{\mu\nu}(x)A_{\alpha\beta}(y):\Big|\,\imath\,\Big\rangle,
\end{equation}
with $\mathfrak{P}^{\mu\nu\alpha\beta}$ being defined as
\begin{equation}\label{4-linear}
\mathfrak{P}^{\mu\nu\alpha\beta}=\overline{\Psi_2}(x)\Gamma^{\mu\nu}\Psi_1(x)\overline{\Psi_3}(y)\Gamma^{\alpha\beta}\Psi_4(y).
\end{equation}
Substituting eq. (\ref{interaction}) and eq. (\ref{psi}), eq. (\ref{4-linear}) becomes
\begin{eqnarray}\label{4-linear1}
\mathfrak{P}^{\mu\nu\alpha\beta} &=&\int\!\!\!\int\!\!\!\int\!\!\!\int\!\!\mathrm{dq_{1}dq_{2}dq_{3}dq_{4}}\Big\{\nonumber\\
& +& c^{\dagger}_{q_2}c_{q_1}d_{q_3}d^{\dagger}_{q_4}[\overline{u}(2)\Gamma^{\mu\nu}u(1)](q,\sigma)[\overline{v}(3)\Gamma^{\alpha\beta}v(4)](q,\sigma)e^{+ix(q_2-q_1)}e^{+iy(q_3-q_4)}\nonumber\\
&+& c^{\dagger}_{q_2}d_{q_1}d_{q_3}c^{\dagger}_{q_4}[\overline{u}(2)\Gamma^{\mu\nu}v(1)](q,\sigma)[\overline{v}(3)\Gamma^{\alpha\beta}u(4)](q,\sigma)e^{+ix(q_2+q_1)}e^{-iy(q_3+q_4)}\nonumber\\
& +& d_{q_2}c_{q_1}c^{\dagger}_{q_3}d^{\dagger}_{q_4}[\overline{v}(2)\Gamma^{\mu\nu}u(1)](q,\sigma)[\overline{u}(3)\Gamma^{\alpha\beta}v(4)](q,\sigma)e^{-ix(q_2+q_1)}e^{+iy(q_3+q_4)}\nonumber\\
& +& d_{q_2}d^{\dagger}_{q_1}c^{\dagger}_{q_3}c^{\dagger}_{q_4}[\overline{v}(2)\Gamma^{\mu\nu}v(1)](q,\sigma)[\overline{u}(3)\Gamma^{\alpha\beta}u(4)](q,\sigma)e^{-ix(q_2-q_1)}e^{+iy(q_3-q_4)}\Big\}.
\end{eqnarray}
Using eq. (\ref{prop}), eq. (\ref{estados0}), eq. (\ref{comutacao0}) and eq. (\ref{4-linear1}) the transition amplitude eq. (\ref{amplitudePAA}) is given by
\begin{eqnarray}\label{Mfinal1}
\mathcal{M} &=&-\frac{\imath}{2\kappa^2}\Big[\,\,\overline{u}(p_3,s_3)\Gamma^\mu_\alpha v(p_4,s_4)\overline{v}(p_2,s_2)\Gamma_\mu^\alpha u(p_1,s_1)\nonumber\\
& +& \overline{u}(p_3,s_3)\Gamma^\beta_\mu v(p_4,s_4)\overline{v}(p_2,s_2)\Gamma_\beta^\mu u(p_1,s_1)\nonumber\\
& -& \overline{u}(p_3,s_3)\Gamma^\mu_\mu v(p_4,s_4)\overline{v}(p_2,s_2)\Gamma_\beta^\beta u(p_1,s_1)\Big],
\end{eqnarray}
where the definition of the four-dimensional delta function, i.e.
\begin{eqnarray}
\int\!\!\!\int\!\!dx\,dy\,e^{-ix(p_1-p_3+\kappa)}e^{-iy(p_2-p_4-\kappa)}=\delta^{(4)}(p_1-p_3+\kappa)\delta^{(4)}(p_2-p_4-\kappa),
\end{eqnarray}
has been used. It is to be noted that the integral of the remaining delta function in eq. (\ref{Mfinal1}) expresses overall four-momentum conservation. By convention, this integral is taken out from the final transition amplitude.

In the cross-section, the square transition amplitude is required, i.e. $|{\cal M}|^2={\cal M}{\cal M}^\dagger$. Then noting that
\begin{equation}
(\,{\Gamma}^{\mu\nu})^\dagger
=\Big( \gamma^0\gamma^{\mu}\gamma^0 p_1^{\nu}+p_2^{\mu}\gamma^0\gamma^{\nu}\gamma^0\Big)
=\gamma^0\Gamma^{\mu\nu}\gamma^0=\overline{\Gamma}^{\mu\nu},
\end{equation}
the $\mathcal{M}^\dagger$ is given as
\begin{eqnarray}\label{Mfinal2}
\mathcal{M}^\dagger &=&\frac{\imath}{2\kappa^2}\Big[\,\,\overline{v}(p_3,s_3)\overline{\Gamma}^\mu_\alpha u(p_4,s_4)\overline{u}(p_2,s_2)\overline{\Gamma}_\mu^\alpha v(p_1,s_1)\nonumber\\
& +& \overline{v}(p_3,s_3)\overline{\Gamma}^\beta_\mu u(p_4,s_4)\overline{u}(p_2,s_2)\overline{\Gamma}_\beta^\mu v(p_1,s_1)\nonumber\\
& -& \overline{v}(p_3,s_3)\overline{\Gamma}^\mu_\mu u(p_4,s_4)\overline{u}(p_2,s_2)\overline{\Gamma}_\beta^\beta v(p_1,s_1)\Big].
\end{eqnarray}
Then the total square transition amplitude becomes
\begin{eqnarray}\label{MM}
|\mathcal{M}|^2 &=& \frac{1}{4\kappa^4}\Big[\,\,\overline{v}(4)\Gamma^\nu_\rho u(3)\overline{u}(1)\Gamma_\nu^\rho v(2)\overline{u}(3)\overline{\Gamma}^\omega_\beta v(4)\overline{v}(2)\overline{\Gamma}^\beta_\omega u(1) \nonumber\\
&+& \overline{v}(4)\Gamma^\nu_\rho u(3)\overline{u}(1)\Gamma_\nu^\rho v(2)\overline{u}(3)\overline{\Gamma}^\omega_\nu v(4)\overline{v}(2)\overline{\Gamma}^\nu_\omega u(1)\nonumber\\
&-& \overline{v}(4)\Gamma^\nu_\rho u(3)\overline{u}(1)\Gamma_\nu^\rho v(2)\overline{u}(3)\overline{\Gamma}^\omega_\omega v(4)\overline{v}(2)\overline{\Gamma}^\eta_\eta u(1)\nonumber\\
&-& \overline{v}(4)\Gamma^\nu_\rho u(3)\overline{u}(1)\Gamma_\nu^\rho v(2)\overline{u}(3)\overline{\Gamma}^\omega_\beta v(4)\overline{v}(2)\overline{\Gamma}^\beta_\omega u(1)\nonumber\\
&+& \overline{v}(4)\Gamma^\nu_\rho u(3)\overline{u}(1)\Gamma_\nu^\rho v(2)\overline{u}(3)\overline{\Gamma}^\omega_\beta v(4)\overline{v}(2)\overline{\Gamma}^\beta_\omega u(1)\nonumber\\
&-& \overline{v}(4)\Gamma^\nu_\rho u(3)\overline{u}(1)\Gamma_\nu^\rho v(2)\overline{u}(3)\overline{\Gamma}^\omega_\omega v(4)\overline{v}(2)\overline{\Gamma}^\eta_\eta u(1)\nonumber\\
&-& \overline{v}(4)\Gamma^\nu_\nu u(3)\overline{u}(1)\Gamma_\rho^\rho v(2)\overline{u}(3)\overline{\Gamma}^\omega_\beta v(4)\overline{v}(2)\overline{\Gamma}^\beta_\omega u(1)\nonumber\\
&-& \overline{v}(4)\Gamma^\nu_\nu u(3)\overline{u}(1)\Gamma_\rho^\rho v(2)\overline{u}(3)\overline{\Gamma}^\omega_\beta v(4)\overline{v}(2)\overline{\Gamma}^\beta_\omega u(1)\nonumber\\
&+& \overline{v}(4)\Gamma^\nu_\nu u(3)\overline{u}(1)\Gamma_\rho^\rho v(2)\overline{u}(3)\overline{\Gamma}^\omega_\omega v(4)\overline{v}(2)\overline{\Gamma}^\eta_\eta u(1)\Big].
\end{eqnarray}

To perform the calculation, the relation
\begin{equation}\label{truque}
     \overline{v}(2)\gamma_\mu p_{1\mu} u(1)\overline{u}(1)\gamma^\mu p_1^{\mu}v(2)=tr\{\gamma_\mu p_{1\mu} u(1)\overline{u}(1)\gamma^\mu p_1^{\mu}v(2)\overline{v}(2)\},
     \end{equation}
and the completeness relations 
     \begin{equation}\label{regras-spinor}
     \sumspin\overline{u}(i)u(i)=\Big(p_i\!\!\!\!\!/\,+m\Big)\quad\mbox{and}\quad \sumspin\overline{v}(i)v(i)=\Big(p_i\!\!\!\!\!/\,-m\Big).
\end{equation}
are used. Thus, calculating the trace and other algebraic calculations, the total transition amplitude is given as
\begin{equation}
\langle|\mathcal{M}|^2\rangle=\frac{1}{4}\sum_{spins}|\mathcal{M}|^2=-\frac{g^4E^4}{512}\Big(3\cos(\theta)-\cos(2\theta)+\cos(3\theta)-3\Big).
\end{equation}
Here, the ultra-relativistic limit has been considered.

Therefore, the gravitational differential cross-section becomes
 \begin{equation}
   \frac{d\sigma}{d\Omega}
   =\frac{1}{64\pi^2\,s}\Big\langle\,\,\Big|\mathcal{M}\Big|^2\,\Big\rangle=-\frac{g^4E^2}{131072\pi^2}\Big(3\cos(\theta)-\cos(2\theta)+\cos(3\theta)-3\Big).
\end{equation}
Performing the integration in angular variables, the cross-section for the gravitational electron-positron scattering process is
\begin{equation}
\sigma_{\!_\mathrm{GEM}}=\frac{\pi G^2E^2}{192c^8}.
\end{equation}
It is interesting to note that although GEM is similar to electromagnetism, this cross-section is different from the quantity obtained in the standard QED, which is given as
\bea
\sigma_{\mathrm{QED}}=\frac{4\pi\alpha^2}{3s},
\eea
where $\alpha$ is the QED coupling constant. Another important difference is that the electromagnetic coupling constant is dimensionless and the GEM coupling constant has dimension, i.e. $g\sim \sqrt{G}$. It is noted that both cross-sections have the same energy dependence, that is, $\sigma\sim\frac{1}{E^2}$. However, at low energy limits the QED cross-section is dominant. On another hand, it is expected that at very high energy limits, where the gravitational effects are strong, the GEM effects will become relevant. Furthermore, it has been investigated whether the gravitational coupling constant can be dimensionless. This is done if a factor involving a characteristic energy scale of the scattering process is added, i.e.  $g\rightarrow g'\sim g E'$. This implies that the gravitational coupling constant increases with the energy. Therefore, at very high energies, the electron-positron scattering process is altered due to gravitational effects.

\section{Conclusion}

Since the beginning of general relativity, it is stated that mass currents must generate a field called the gravitomagnetic field, in analogy with electromagnetism. The effects of gravity associated with the rotation of massive bodies can be better understood using a formal analogy with electromagnetism. In this context, the GEM theory and its Lagrangian have been considered. The gravitational cross-section for the electron-positron scattering process with gravitons as an intermediate state has been calculated. The corresponding effects are expected to be small, but can be measurable on a very high energy scale where the gravitational interaction is dominant. The gravitational and electromagnetic cross sections are compared. It is observed that the electromagnetic coupling constant is dimensionless while the gravitational coupling constant depends on the gravitational constant $G$. On the other hand, both cross-sections have the same energy dependence. Our results show that a well-known QED scattering process may be corrected by gravitational effects. Furthermore, successful experimental results, such as those obtained by the Gravity Probe-B experiment, among others, indicate that the contribution of the GEM is relevant for scattering processes like the one studied here and can be measured in future experiments.

\section*{Acknowledgments}

This work by A. F. S. is supported by National Council for Scientific and Technological Develo\-pment - CNPq projects 430194/2018-8 and 313400/2020-2.

\end{document}